\documentclass[traditabstract]{aa}

\usepackage{txfonts}
\usepackage{graphicx}
\usepackage{natbib}
\usepackage{longtable}

\def\spose#1{\hbox to 0pt{#1\hss}}
\def\simlt{\mathrel{\spose{\lower 3pt\hbox{$\mathchar"218$}}
     \raise 2.0pt\hbox{$\mathchar"13C$}}}
\def\simgt{\mathrel{\spose{\lower 3pt\hbox{$\mathchar"218$}}
     \raise 2.0pt\hbox{$\mathchar"13E$}}}

\def\farcmin{\ifmmode \rlap.{^{\prime}}\else
 Ê Ê$\rlap.{^{\prime}}$\fi}

\begin{document}

\title{Spectroscopic confirmation of a galaxy cluster\\ associated with 7C~1756+6520 at $z=1.416$}

\author{Audrey Galametz\inst{1,2}
\and Daniel Stern\inst{3}
\and S. Adam Stanford\inst{4}
\and Carlos De Breuck\inst{1}\and \\
Jo\"{e}l Vernet\inst{1} 
\and Roger L. Griffith\inst{3}
\and Fiona A. Harrison\inst{5}}

\institute{European Southern Observatory, Karl-Schwarzschild-Stra$\betaup$e 2, D-85748 Garching, Germany [e-mail: {\tt agalamet@eso.org}]
\and Observatoire Astronomique de Strasbourg, 11 rue de l$'$Universit\'e, 67000 Strasbourg, France
\and Jet Propulsion Laboratory, California Institute of Technology, 4800 Oak Grove Dr., Pasadena, CA 91109
\and Institute of Geophysics and Planetary Physics, Lawrence Livermore National Laboratory, Livermore, CA 94550
\and Space Radiation Laboratory, MS 220-47, California Institute of Technology, Pasadena, CA 91125
}

\abstract{
We present spectroscopic follow-up of an
overdensity of galaxies photometrically selected to be at $1.4 < z <
2.5$ found in the vicinity of the radio galaxy 7C~1756+6520 at $z =
1.4156$.  Using the DEIMOS optical multi-object spectrograph on the
Keck 2 telescope, we observed a total of $129$ $BzK$-selected sources, comprising $82$
blue, star-forming galaxy candidates (s$BzK$) and $47$ red,
passively-evolving galaxy candidates (p$BzK$*), as well as $11$
mid-infrared selected AGN candidates. We obtain robust spectroscopic
redshifts for $36$ blue galaxies, $7$ red galaxies and $9$ AGN
candidates.  Assuming all 
foreground interlopers were identified, we find that only $16$\% ($9$\%) of the
s$BzK$ (p$BzK$*) galaxies are at $z<1.4$.  Therefore, the $BzK$
criteria are shown to be relatively robust at identifying galaxies
at moderate redshifts.  Twenty-one galaxies, including the radio
galaxy, four additional AGN candidates and three red galaxy candidates
are found with $1.4156 \pm 0.025$, forming a large scale structure
at the redshift of the radio galaxy.  Of these, eight have projected
offsets $<2$~Mpc relative to the radio galaxy position and have
velocity offsets $<1000$~km~s$^{-1}$ relative to the radio galaxy
redshift.  This confirms that 7C~1756+6520 is associated
with a high-redshift galaxy cluster.  A second compact group of
four galaxies is found at $z\sim1.437$, forming a sub-group offset by 
$\Delta v\sim3000$~km/s and approximately $1\farcmin5$ east of the radio galaxy.}


\maketitle

\section{Introduction}

Galaxy clusters provide an important tool for studying both the
formation of galaxies and for deriving cosmological parameters.
Out to the furthest redshifts studied thus far, the oldest, most massive
galaxies reside within clusters, and thus clusters provide ideal
probes of the formation and evolution of galaxies.  As
the largest collapsed structures in the universe, the cosmic history
of galaxy clusters is sensitive to key cosmological parameters
\citep[e.g.,][]{Vikhlinin2009, Stern2010}.  Significant work has
been done to discover more distant galaxy clusters.  However,
the number of clusters currently confirmed at $z>1$ remains low.
Two of the most distant confirmed galaxy clusters were identified
from the extended X-ray emission of the intracluster medium:
XMMU~J2235.3-2557 at $z=1.39$ \citep{Mullis2005, Lidman2008,
Rosati2009} and XMMXCS~J2215.9-1738 at $z=1.46$ \citep{Stanford2006,
Hilton2007}.  However, the X-ray identification of candidate 
clusters is very difficult at $z>1$ since the surface brightness
of the extended X-ray emission fades as $(1+z)^4$.

Another method to find galaxy clusters has been to detect overdensities
of red sources in optical imaging data using the so-called `red
sequence method'. \citet{Gladders2000} showed that two filter imaging
is sufficient to perform a cluster search through the detection of
the red sequence of early-type galaxies.  The colours of such
galaxies are quite distinct due to the strong $4000$\AA\ break
(D4000) in their spectra.  However, this break shifts into the
near-infrared at $z>1.5$, and the colors can become degenerate.  Using sensitive mid-infrared data obtained
with the {\it Spitzer Space Telescope}, the {\it Spitzer} Adaptation
of the Red-Sequence Cluster Survey \citep[SpARCS;~][]{Wilson2009}
team have recently pushed the technique to higher redshift and
confirmed one galaxy cluster at $z = 1.34$.   Using full 
photometric redshifts with effectively a stellar mass-selected galaxy sample,
\citet{Eisenhardt2008} have identified 106 galaxy cluster candidates
at $z > 1$, 13 of which have been spectroscopically confirmed to date.  This
technique, which does not depend on the presence of a red sequence, has confirmed three galaxy clusters at $z > 1.3$,
with the most distant at $z = 1.41$ \citep{Stanford2005}.

An alternative method to find high-redshift galaxy clusters is to
look in the surroundings of powerful, high-redshift radio galaxies
(HzRGs).  HzRGs are among the most massive galaxies in the Universe
\citep[$M_{\rm star}>10^{11}~M_\odot$;][]{Seymour2007} and therefore
are likely to inhabit dense regions.  Narrow-band imaging surveys have been
intensively conducted in the surroundings of HzRGs, mostly at $z>2$
to search for overdensities of Ly$\alpha$ and/or H$\alpha$ emitters.
Numerous overdensities have been detected and spectroscopically
confirmed around radio galaxies at $z>2$, reaching even to $z=5.2$
\citep[TN~J0924-2201;~][]{Venemans2007}. However, Ly$\alpha$ emitters
are small, faint, young star-forming galaxies with masses of a few
$\times$ $10^8 M_{\odot}$ \citep{Overzier2008}, and probably
represent only a small fraction of the total stellar mass of these clusters.
Furthermore, at such high redshifts, these overdensities are suspected
to still be forming and not yet bound. The term `protoclusters' is
commonly used to describe such systems. A complementary approach is
to isolate the evolved, massive galaxy population near HzRGs using
purely near-infrared colour selection \citep{Kajisawa2006, Kodama2007}.
Though this method has successfully found overdensities of red
galaxies at $z \sim 2$, it has been challenging to spectroscopically
confirm their association with the HzRGs \citep[e.g.,][]{Doherty2010}.
A few studies have also applied related methods to slightly lower
redshift HzRGs --- e.g., \citet{Stern2003} and \citet{Best2003}
found overdensities of extremely red galaxies in the environments
of radio-loud active galactic nuclei (AGN) at $z \sim 1.5$.

Recently, \citet{Galametz2009B} (G09 hereafter) presented an
overdensity of galaxy candidates at $z>1.4$ in the field of the
radio galaxy 7C~1756+6520. The radio galaxy was initially reported 
to be at $z = 1.48$ by \citet{Lacy1999} based on the tentative
identification of a single, uncertain emission feature.  Based on
deeper Keck spectroscopy, we find that the radio galaxy is, in fact,
at $z=1.4156$ (see \S3.1). G09 made use of a revised version of the
so called `$BzK$ criteria', a  two-colour selection technique based
on $BzK$ photometry \citep{Daddi2004} to isolate galaxies at $1.4
\le z \le 2.5$ and classify them as either red, passively evolving
(p$BzK$) or blue, star-forming (s$BzK$) systems. The star-forming
candidates are selected by $BzK \equiv (z - K) - (B - z) > -0.2$.
The original \citet{Daddi2004} criterion selected passive (p$BzK$)
systems by $BzK < -0.2 \cap (z - K) > 2.5$, but was empirically
shown in G09 to have a low success rate at $z \sim 1.4$.  In G09,
we therefore extended the selection criteria to reliably identify
galaxies at the low-redshift end of the $BzK$ criteria by defining
p$BzK$* galaxies to have $BzK < -0.2 \cap (z - K) > 2.2$.

G09 selected s$BzK$ and p$BzK$* galaxies in the field around
7C~1756+6520 using deep, multiwavelength data: $B$-band and $z$-band
images from the Large Format Camera \citep[LFC;][]{Simcoe2000} on
the Palomar 5m Hale telescope and a $Ks$-band (hereafter $K$) image
from the Wide-field Infrared Camera \citep[WIRCAM;][]{Puget2004}
on the Canada-France-Hawaii telescope (see G09, Table~1 for details).

Relative to the four deep MUSYC blank fields \citep{Gawiser2006,
Quadri2007}, G09 found an overdensity of cluster member candidates
around 7C~1756+6520 for both s$BzK$ and p$BzK$* galaxies --- by a
factor of $2$ and $4.7$, respectively.  Using the {\it Spitzer}/IRAC
colour-colour selection of \citet{Stern2005}, we also isolated $12$
mid-infrared selected AGN candidates around 7C~1756+6520, which
represents an overdensity by a factor of two compared to the IRAC
Shallow Survey \citep{Eisenhardt2004}.

We describe in this paper the results of our spectroscopic follow-up
of the overdensity found in the surroundings of 7C~1756+6520. The
next section presents the selected targets, observations and data
reduction. Section 3 describes the results of our spectroscopy and
presents the redshift distribution of the observed sources. Section
4 reports the discovery of a concentration of $20$ galaxies whose
redshifts are close to that of the radio galaxy.  We assume a
$\Lambda$CDM cosmology with $H_0 = 70$ km s$^{-1}$ Mpc$^{-1}$,
$\Omega_m = 0.3$ and $\Omega_{\Lambda} = 0.7$. The magnitudes are
expressed in the AB photometric system unless otherwise stated.

\section{Keck/DEIMOS Spectroscopy}

In order to examine whether the detected overdensities were indeed
associated with 7C~1756+6520, we started a spectroscopic follow up
campaign at the Keck~2 telescope using the Deep Imaging Multi-Object
Spectrograph \citep[DEIMOS;~][]{Faber2003}.

We observed one slit mask on UT~2008 August 31 using the 600ZD
grating which is blazed at $7500$~\AA, covers a typical wavelength
range of $5000-10000$~\AA, and has a spectral resolution of $3.7$~\AA\
(FWHM).  This slitmask targeted $47$ sources, including one AGN
candidate, $22$ s$BzK$ galaxies and $24$ p$BzK$* galaxies.  Eight
1800s exposures were obtained over the course of three nights using
the same mask; conditions were clear with 0\farcs8 seeing on average.

Two additional slit masks were observed in September 2009, again
with DEIMOS and the 600ZD grating.  The first mask, observed for a
total of 2.5~hr on UT~2009 September 16, contained $54$ objects:
the radio galaxy itself, five AGN candidates, $34$ s$BzK$ galaxies
and $14$ p$BzK$* galaxies.  This slit mask was oriented so that
a p$BzK$* galaxy $3\arcsec$ south-west of 7C~1756+6520 was on the
same slitlet as the radio galaxy itself (e.g., PA$=-61.57\deg$).
The second slit mask, observed for a total of 1.75~hr on UT~2009
September 17, contained $56$ sources: seven AGN candidates, $36$
s$BzK$ galaxies and $13$ p$BzK$* galaxies.  Both nights were
exceptional, with photometric conditions and 0\farcs5 seeing.

\begin{figure}
\begin{center} 
\includegraphics[width=6.8cm,angle=0,bb=50 20 550 440]{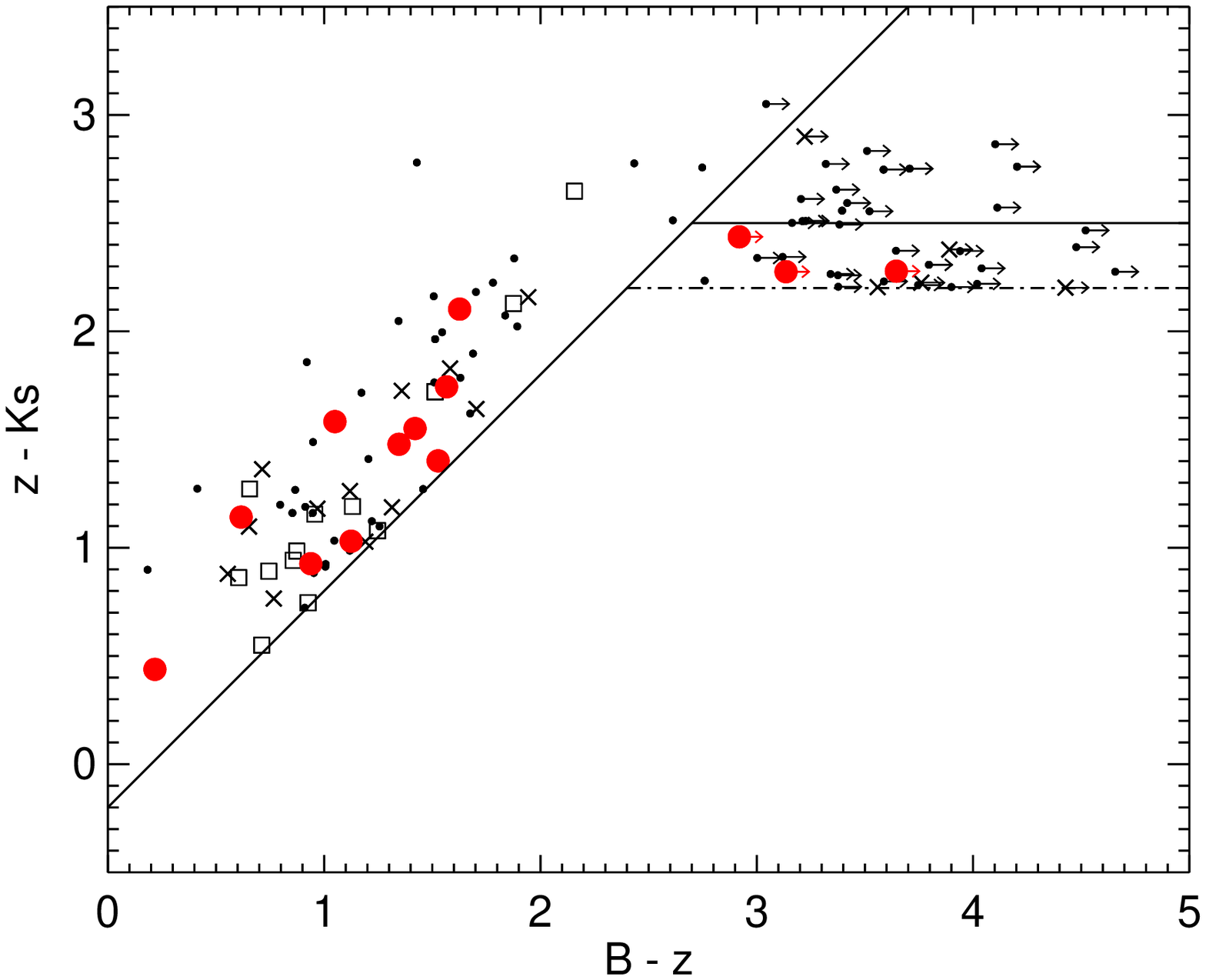}\\
\includegraphics[width=6.8cm,angle=0,bb=50 20 550 400]{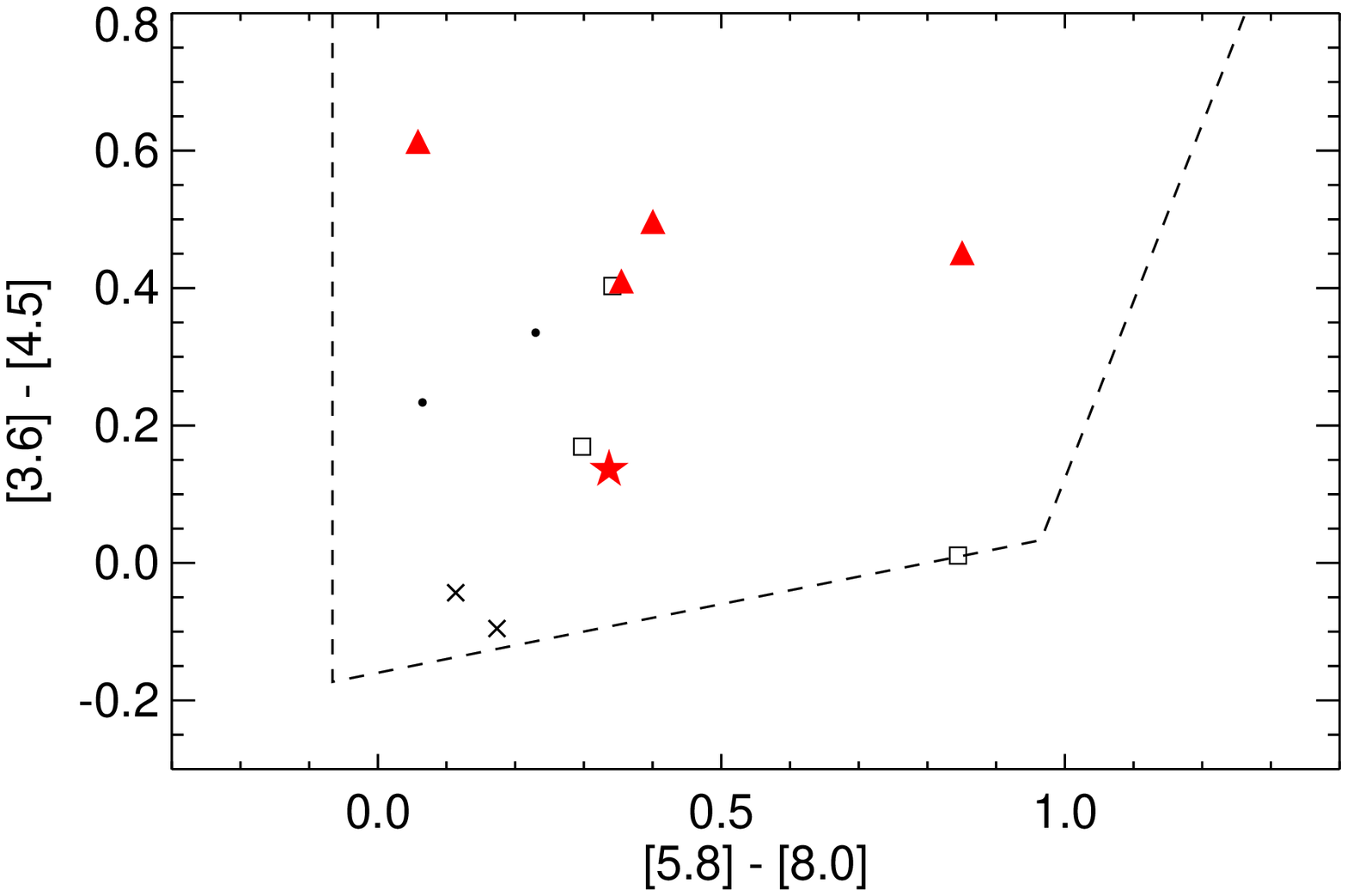}
\includegraphics[width=6.8cm,angle=0,bb=50 20 550 600]{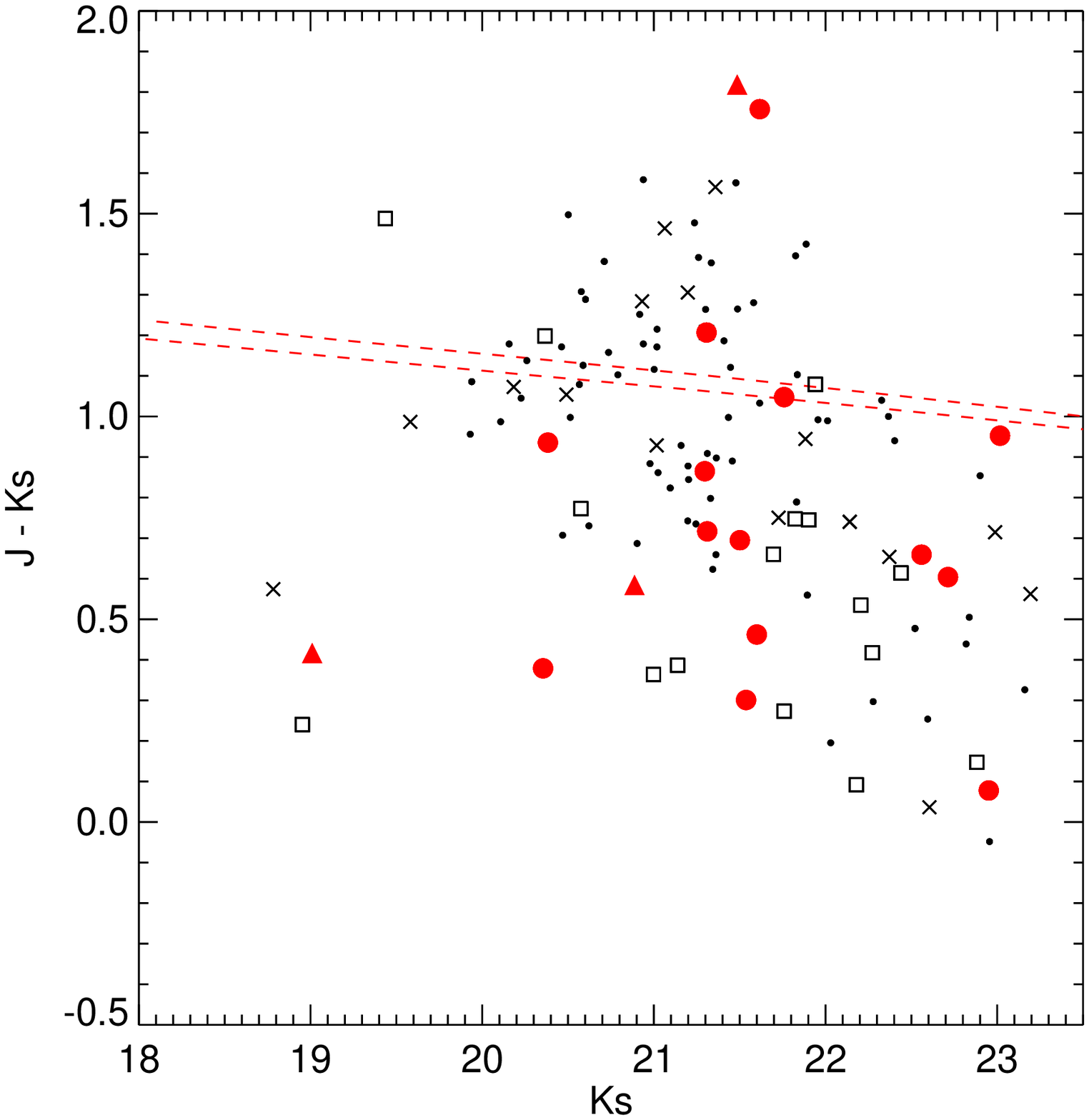}
\end{center}
\caption[Our spectroscopic targets.]
{Keck/DEIMOS spectroscopic targets. {\it Top: }Distribution of
targeted $BzK$ galaxies in the modified \citet{Daddi2004} `$BzK$'
diagram. {\it Middle: }Distribution of targeted AGN candidates and
7C~1756+6520 (red star) in the IRAC colour-colour diagram from
\citet{Stern2005}. {\it Bottom: }Colour-magnitude diagram ($J-K$
vs. $K$) of spectroscopic targets with reliable $J$ and/or $K$
photometry. We also plot the expected locations of the red sequence
at $z=1.42$ (dashed line) for
a formation redshifts of $z_f=4$
(lower curve) and $z_f=5$ (upper curve).  For the three panels,
small dots indicate all sources observed but not yielding a redshift.
Spectroscopically confirmed members of the structure around
7C~1756+6520 (e.g., with $\mid z-z_{\rm HzRG}\mid\ < 0.03$) are
indicated by red symbols (points for $BzK$-selected galaxies and
triangles for AGN candidates). Other sources with spectroscopic
redshifts are indicated by crosses and squares for background and
foreground sources, respectively.}
\label{targets}
\end{figure}

\begin{figure}
\begin{center} 
\includegraphics[width=8.5cm,angle=0]{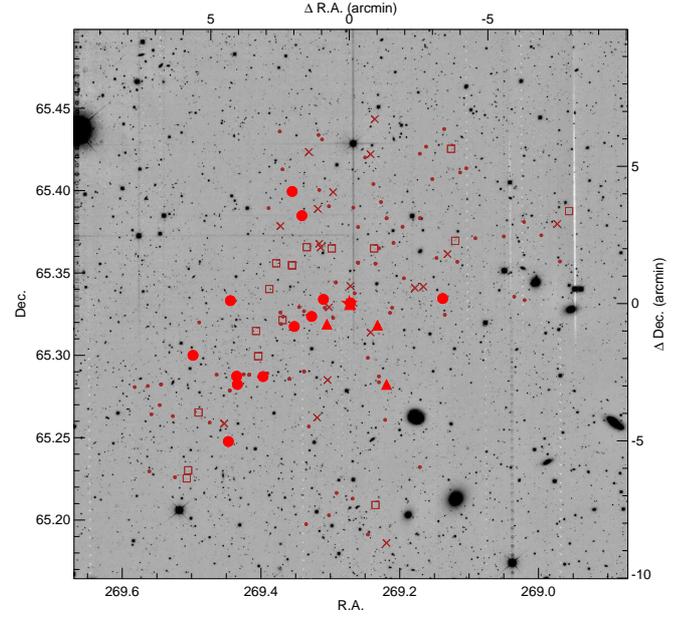}
\end{center}
\caption[CFHT/WIRCam $J$-band image of 7C1756+6520 illustrating the
spatial distribution of our targets.]
{$J$-band image of 7C1756+6520 (red star) illustrating the
spatial distribution of our targets. See Fig.~\ref{targets} for
description of symbols. Left and bottom axes are in J2000 coordinate
system; right and top axes indicate distances relative to the radio
galaxy.  Image is $20\arcmin \times 20\arcmin$ and was obtained
with WIRCAM on CFHT.} 
\label{radec}
\end{figure}

Fourteen $BzK$ galaxies and two AGN were observed multiple times
across these three slit masks.  Thus, the total number of distinct
targets was $129$ $BzK$-selected targets --- including $82$ ($47$)
s$BzK$ (p$BzK$*) galaxies --- as well as $11$ AGN candidates and
the radio galaxy itself.  Fig.~\ref{targets} shows the distribution
of the targets in the two colour-colour diagrams used in G09 to
select candidate cluster members: the `$BzK$' diagram for  the
$BzK$-selected galaxies and the IRAC $[3.6]-[4.5]$ vs. $[5.8]-[8.0]$
diagram for the mid-infrared selected AGN. We refer to G09 for details
on the selection techniques.  The last panel of Fig.~\ref{targets}
shows the distribution of our targets in a $J-K$ vs.  $K$
colour-magnitude diagram. Fig.~\ref{radec} shows the spatial
distribution of these targets around the radio galaxy. The targeted
sources have optical magnitudes in the range $20.8<z<24.5$ with
$\langle z \rangle \sim 23.6$ and near-infrared magnitudes in the
range $20.2<K<23.3$ with $\langle K \rangle \sim 21.6$.

All data were processed using a slightly modified version
of the pipeline developed by the DEEP2 team at UC-Berkeley, and
data were flux calibrated using archival sensitivity functions for
the same instrument configuration derived using standard stars from
\citet{Massey1990}. The uncertainty on the wavelength calibration,
derived from sky lines, was found to be $0.4$\AA~and consistent from mask to mask.

\section{Results}

\subsection{Spectrum of 7C~1756+6520}

\begin{figure}
\begin{center} 
\includegraphics[width=6cm,angle=-90]{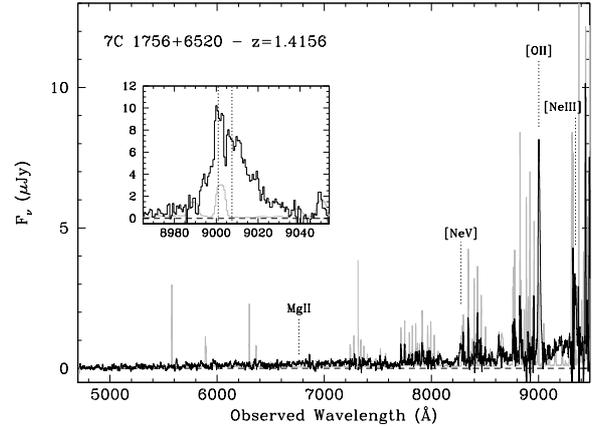}
\end{center}
\caption[Keck/DEIMOS spectrum of 7C~1756+6520 obtained in September
2009.] {Keck/DEIMOS spectrum of 7C~1756+6520 obtained on UT September
16 (black, with a smoothing box of 13~\AA\ applied for clarity
in the large panel; the inset, detailing the [OII] doublet, was not
smoothed).  The sky spectrum, in grey, has been scaled down by a
factor of 10.  The locations of prominent lines are labeled (MgII
is not detected in these data).}

\label{HzRG}
\end{figure}

\citet{Lacy1999} tentatively assigned $z=1.48$ to the radio galaxy
based on an uncertain emission line at 6005~\AA\ assumed to be
[NeIV]$\lambda2425$\AA.  This redshift was assigned a quality
`$\gamma$', indicating an `uncertain' redshift.  We obtained a quick
($10$~min) spectrum of the radio galaxy during twilight on UT~2009
April 27 using the Low Resolution Imaging Spectrometer
\citep[LRIS;~][]{Oke1995} on the Keck~1 telescope.  These data
detected only a single, faint emission line at 6759~\AA\ from 
which no reliable redshift could be assigned.  The
deeper Keck/DEIMOS spectrum obtained in September 2009 (Fig.~\ref{HzRG})
yielded a redshift of $1.4156$ based on three clearly detected
emission lines: [NeV]$\lambda3426$\AA, [OII]$\lambda3727$\AA~and
[NeIII]$\lambda3869$\AA.  The [OII] line (Fig.~\ref{HzRG}, inset)
is clearly split in these data.  In retrospect, the faint emission line in the LRIS
spectrum in fact corresponds to MgII$\lambda2800$\AA.  We note that
the emission line in \citet{Lacy1999}, from which their redshift
identification was based, is not detected in either of our deeper
Keck spectra.  It would correspond to restframe $2486$~\AA~at the
redshift of the radio galaxy and was therefore most likely a spurious
detection.

\subsection{Spectroscopy of candidate cluster members}

The redshift range $1.4<z<2.5$ is often described as the `redshift
desert' because optical spectroscopic confirmation of targets in
that redshift range is challenging, with most of the strongest
classical spectral features  (e.g.,
[OII]$\lambda3727$\AA, D$4000$, H$\beta\lambda4861$\AA,
[OIII]$\lambda5007$\AA, H$\alpha\lambda6563$\AA) redshifted
longward of the wavelength range where CCDs are most sensitive
(e.g., $4000 - 9000$~\AA).  DEIMOS uses a modern, red-sensitive CCD
detector which, combined with the grating we had installed for these
observations, provides data out to $\sim 1~\mu$m and would, in
principle, detect the [OII] emission line to $z \sim 1.7$.  However,
the sensitivity of the spectrograph detectors rapidly decreases at
the longest wavelengths and telluric OH emission becomes progressively
problematic, especially at $\lambda>9300$~\AA~ (corresponding to
$z>1.5$ for the [OII] line).

A visual inspection of the reduced spectra permitted us to assign
spectroscopic redshifts ($z_{\rm spec}$) to $43$ $BzK$ galaxies,
$9$ AGN candidates and the radio galaxy (see \S 3.1).  Due to the
design of the masks and the length of the slits, we also obtained
additional spectroscopic redshifts of several serendipitous sources.
We determined redshifts by fitting emission lines by gaussian
profiles. When clearly split, the [OII] doublet was fit by a double
gaussian profile. We derived uncertainties on the spectroscopic redshifts
by adding in quadrature the fitting uncertainties and the small uncertainty in
the wavelength calibration (see \S 2; $0.4$\AA~corresponds to $\Delta z\sim0.0001$).
We assigned a quality flag `A' or `B' to all measured redshifts: `A'
indicates a highly certain $z_{\rm spec}$ based on at least two
spectral features or [OII]$\lambda3727$\AA~being clearly identified
as a doublet; `B' indicates high-level confidence in the $z_{\rm
spec}$ based on one well-detected spectral feature. Lower confidence
quality flags were recorded during our data analysis, but are not
reported here. 

\begin{figure}[!t]
\begin{center} 
\includegraphics[width=9cm,angle=0]{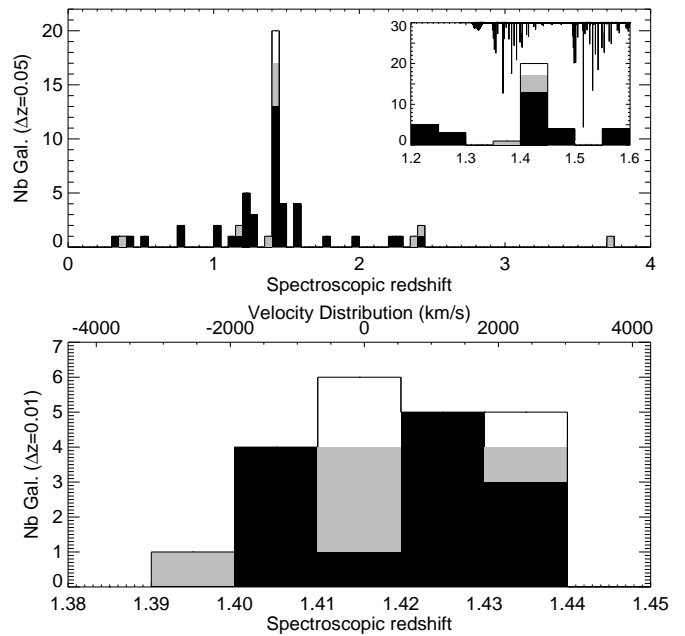}
\end{center}
\caption[Histogram of the $52$ assigned spectroscopic redshifts.]
{Histogram of the $52$ spectroscopic redshifts, including the $43$
$BzK$-selected objects (black), the nine mid-infrared selected
AGNs and the radio galaxy (the AGN are all shown in grey).  We also
include the spectroscopic redshifts of the three serendipitous
objects detected in our slits with $\mid z-z_{\rm HzRG}\mid\ <0.025$
(white; see \S4.2). The inset shows the $1.2<z<1.6$ redshift range
with sky lines overlaid from the top axis.  The bottom panel shows
the spectroscopic redshift distribution near the redshift of
7C~1756+6520, with the corresponding velocity distribution relative
to $z=1.4156$ on the top axis.}
\label{specz}
\end{figure}

Fig.~\ref{specz} presents the distribution of our spectroscopic
redshifts.  Coordinates, $z_{\rm spec}$, redshift, quality flags
(Q), magnitudes and the target selection criteria of all confirmed
sources are provided in Table~\ref{members} (for members of the
HzRG large-scale structure) and Table~\ref{others} (for the
foreground/background sources). The final column of these tables
lists the spectral features used to determine redshifts.

\subsubsection{$BzK$-selected sources}

We obtained spectroscopic redshifts for $43$ ($33$\%) of the targeted
$BzK$-selected sources, including $36$ ($44$\%) of the s$BzK$ targets
and seven ($15$\%) of the p$BzK$* targets. The majority ($79$\%)
of the redshifts were calculated from the [OII]$\lambda3727$\AA~doublet,
which is resolved at the spectral resolution of the DEIMOS data.
Other features, such as MgII$\lambda2798$\AA\ (either in absorption
or emission) or the D4000 break, were also present in several of
the spectra, permitting us to unambiguously confirm our redshift
identifications.


Of the sources for which we derived successful spectroscopic
redshifts, $64$\% ($43$\%) of the s$BzK$ (p$BzK$*) galaxies are at
$z_{\rm spec}>1.4$.  Based on these spectroscopic results, it is
challenging to derive definitive conclusions regarding the reliability
of the $BzK$ criterion to select high-redshift galaxies.  At $z>1.3$,
bright and numerous sky lines (see Fig.~\ref{specz}, sub-panel)
coupled with decreasing detector sensitivity at longer wavelengths
limits our ability to measure redshifts based on the [OII] doublet.
Furthermore, our redshift coverage does not permit detection of
[OII] at $z>1.7$ (and at slightly lower redshifts for many sources
since the spectral coverage of any given source will depend on where
it is located on the slitmask).  With these caveats in mind, we now
briefly discuss the efficiency of the two $BzK$ criteria to identify
galaxies at $z > 1.4$.

{\it s$BzK$ galaxies:}  Assuming that the s$BzK$ criterion successfully
identifies sources with strong signatures of star formation ---
e.g., strong [OII], [OIII] and H$\alpha$ emission lines --- then
our Keck/DEIMOS data should have yielded robust spectroscopic
redshifts for the vast majority of s$BzK$ galaxies at $z<1.4$.
Indeed, all but four of the successfully identified s$BzK$ sources
show these emission lines.  The last four are at $1.6 < z <
2.4$ where the [OII] line had shifted beyond the spectroscopic
coverage of our instrument; their redshifts instead are based on 
absorption lines such as CIV$\lambda$1549 and
AlII$\lambda$1670 which are commonly seen in the UV spectra of
star-forming galaxies. If we thus assume
that our data identified all of the interlopers and that the failed
sources are all at $z > 1.4$, then we find that the s$BzK$ criterion
is quite robust, with only $16$\% of such targets at $z<1.4$.

{\it p$BzK$* galaxies:} As expected, the p$BzK$* galaxies often
presented very red optical continuum with no clear features (e.g.,
spectral breaks) --- indeed, all seven of the
p$BzK$* galaxies for which we did derive spectroscopic redshifts
had [OII] emission features.  For evolved galaxies, the D4000 break
is generally the strongest spectral signature.  However, assuming
that a spectrum must cover out to restframe $4100$~\AA\ to robustly
identify this feature, then the $1~\mu$m cut-off of the DEIMOS data
implies that we can only identify D4000 out to $z = 1.44$; we are
thus forced to rely on weaker spectroscopic features such as B2900 or B2640
to identify evolved galaxies at $z \sim 1.5$ and beyond.  This makes
redshift determinations challenging, and is presumably the
reason for our low success rate for the p$BzK$* sources.  Assuming
that our data identified all of the foreground interlopers and that
the failed sources are all at $z > 1.4$, we find that the
p$BzK$* criteria are also robust, with only 9\% of such targets
at $z < 1.4$.  We note, however, that this assumption is less robust
than the similar assumption for the s$BzK$ galaxies, given the low signal-to-noise
ratio of the p$BzK$ spectra.

\begin{figure*}[!t] 
\begin{center} 
\includegraphics[width=10cm,angle=0,bb=100 120 500 500]{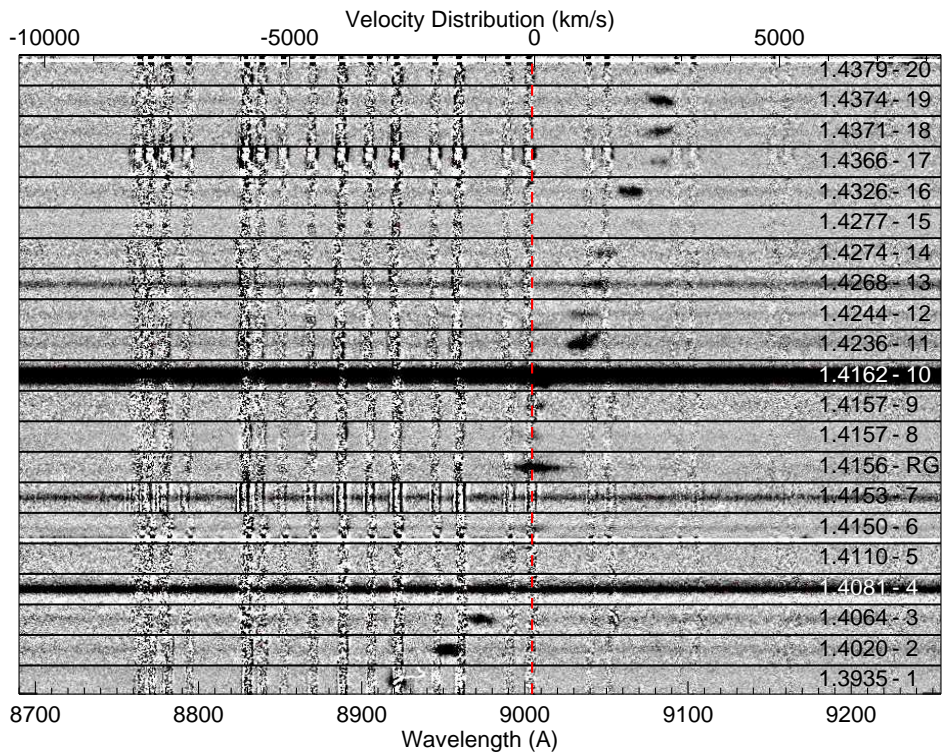}
\end{center}
\caption[Two-dimensional spectra of the spectroscopically confirmed members of the structure.]
{Two-dimensional spectra of the spectroscopically confirmed members
of the structure, centered near the [OII] emission line.  The
position of the [OII] doublet at $z=1.4156$ is indicated by the
vertical red dashed line.  The redshift and ID number of each source
is provided on the right end side. The top axis indicates the
corresponding velocity distribution relative to the redshift of the
radio galaxy.}
\label{2D}
\end{figure*}

In the colour-magnitude diagram in Fig.~\ref{targets} (bottom panel),
we show the expected location of the red sequence at $z=1.42$ for
a formation redshift $z_f=4-5$ \citep[T. Kodama, private communication;
see also~][]{Kodama1998}.  We targeted numerous p$BzK$* galaxies
with colours consistent with the red sequence for which we unfortunately
could not determine a redshift.  However, the three
p$BzK$* galaxies that were confirmed to be at $z_{\rm spec}>1.4$
all have $2.2<z-K<2.5$ would have been missed by the standard
\citet{Daddi2004} criterion for p$BzK$ galaxies (see \S1). Our
modified p$BzK$* criterion presented in G09 increases the completeness
of galaxies at $z \sim 1.4$.  On the other hand, our selection is also
contaminated by lower redshift (albeit still relatively distant)
objects. The other four p$BzK$* with $z_{\rm spec}$ are foreground
interlopers with $1.18<z<1.26$.



\subsubsection{Mid-infrared selected AGN candidates}

Eleven of the $12$ AGN candidates presented in G09 were observed
and we were able to assign spectroscopic redshifts for $9$ of these $11$.
The AGN selection criterion of \citet{Stern2005} does not preferentially
select AGN in any specific redshift range (other than an inefficiency
at selecting sources at $z \sim 4$ where H$\alpha$ shifts into the
IRAC channel~1 and produces blue [3.6]$-$[4.5] colours;~ see Assef
et al.~2010)\nocite{Assef2010}.  The mid-infrared selection criteria
are also sensitive to both obscured and unobscured AGN
\citep[e.g.,][]{Eckart2010}.  Unsurpringly, therefore, we identified
AGN at a range of redshifts ($0.37 < z < 3.70$) and with a range
of properties.

The three AGN at $z>2$ are all classical (e.g., unobscured or type~1)
quasars showing broad CIV$\lambda1549$\AA~and
CIII]$\lambda1909$\AA~emission lines.  Two AGN candidates
(Table~\ref{members}, sources $1$ and $18$) do not show any classical
AGN features.  The first was assigned a quality B spectroscopic
redshift of $1.3935$ based on a single emission line (which falls
on a sky line) attributed to [OII], though the spectrum 
also shows hints of spectral breaks at restframe $2900$~\AA~and
$3260$~\AA, making the inferred redshift quite likely.  The second
source shows a resolved [OII] doublet at $z = 1.4371$.  Two AGN
candidates at $z_{\rm spec}>1$ show both the narrow [OII] doublet
(which could be due to AGN activity and/or star formation) as well
as broad MgII emission, which clearly indicates an AGN.  Four of
the mid-infrared selected AGN candidates sources are at redshifts
close to the radio galaxy and are discussed more in \S4.2 (see
Table~\ref{members}).

\section{A structure of galaxies associated with 7C~1756+6520}

\subsection{Companions close to the radio galaxy}

In G09 we stated that an s$BzK$ galaxy, a p$BzK$* galaxy and an AGN
candidate are all found within $6\arcsec$ of 7C~1756+6520.  For $z
= 1.4156$, this corresponds to a projected separation of 50~kpc.
The probability of finding three candidates in such a small area
is less than $0.2$\%, suggesting that these sources are all associated
with the radio galaxy.  Both the p$BzK$* and the AGN candidates
were observed with DEIMOS. The p$BzK$* source shows a clear, resolved
[OII] doublet at $z=1.4244$, and is thus associated with the radio
galaxy (Table~\ref{members}, ID $12$; quality A). The AGN candidate
shows a single, broad emission line which we interpret to be MgII
at $z=1.4153$ (Table~\ref{members}, ID $7$; quality B). The s$BzK$ galaxy was
not observed. We therefore believe that the radio galaxy has 
at least one, most probably two, and perhaps three close-by companions.


\subsection{Two compact sub-groups in a large scale structure}

A clear peak is seen in the spectroscopic redshift distribution
near the radio galaxy redshift (Fig.~\ref{specz}).  Not including
the HzRG, $20$ galaxies are found with $\mid z-z_{\rm HzRG}\mid\,
<0.025$, corresponding to peculiar velocities $\simlt 3000$\, km\,
s$^{-1}$ with respect to the radio galaxy.  Table~\ref{members}
reports the coordinates and $BzK$ magnitudes of these $20$ galaxies
($21$ including the HzRG), comprising four mid-infrared AGN candidates,
three p$BzK$* galaxies, $10$ s$BzK$ galaxies and three serendipitous
galaxies.  The first serendipitous galaxy, Cl1756.6 (or serendip.1),
is blended in all of our imaging bands with an object at $z=0.76$.
The foreground object, whose photometry is contaminated in all three
bands by its projected neighbor, was one of our targeted s$BzK$
galaxies.  The second (Cl1756.8, or serendip.2) was observed in the
same slit as Cl1756.9, one of the s$BzK$ targets. The two sources
are close both on the plane of the sky and in redshift, and are
therefore close-by companions. Cl1756.8 is very faint at $K$ (detected
at $<2\sigma$) and was not considered by our $BzK$ criteria.  The
last serendipitous object (Cl1756.20, or serendip.3) was not selected
by any of our criteria.  The coordinates and magnitudes of the three
p$BzK$* (Cl1756.5, Cl1756.12 and Cl1756.15 in Table~\ref{members})
were previously reported in G09 as galaxies $31$, $43$ and $63$,
respectively (see Table~5 of G09). G09 did not tabulate the s$BzK$
candidates.


The two dimensional spectra of the $20$ galaxies associated with
7C~1756+6520 are shown in Fig.~\ref{2D}, centered near the [OII]
line used to determine their redshifts.  As expected, the p$BzK$*
galaxies (sources $5$, $12$ and $15$) show fainter [OII] emission
lines than most of the AGN or star-forming (s$BzK$) candidates.

There is still no clear and agreed upon definition of 
a galaxy cluster. \citet{Eisenhardt2008} considers a $z>1$ cluster
spectroscopically confirmed if five galaxies are robustly identified
within a radius of $2$~Mpc and $\pm2000\, (1+\langle z_{\rm
spec}\rangle)$\, km\, s$^{-1}$.  
According to this definition, $14$ objects
(including the radio galaxy) would therefore be part of our galaxy
cluster.  However, the spectroscopically-confirmed, high-redshift
galaxy clusters in \citet{Eisenhardt2008} and elsewhere usually show much narrower
velocity distributions.  The spectroscopically confirmed members
of XMMXCS~J2215.9-1738 at $z=1.457$ all have peculiar velocities
offset by less than $1000$\, km\, s$^{-1}$ with respect to the
cluster redshift \citep{Hilton2007}. Recently,
\citet{Hayashi2009} presented an [OII] emission survey of this same
cluster.  They identified $44$ [OII] emitters over a larger area centered on the cluster with
$\Delta v<2000$\, km\, s$^{-1}$ with respect to the cluster redshift.

\begin{figure*}
\begin{center} 
\includegraphics[width=14cm,angle=0]{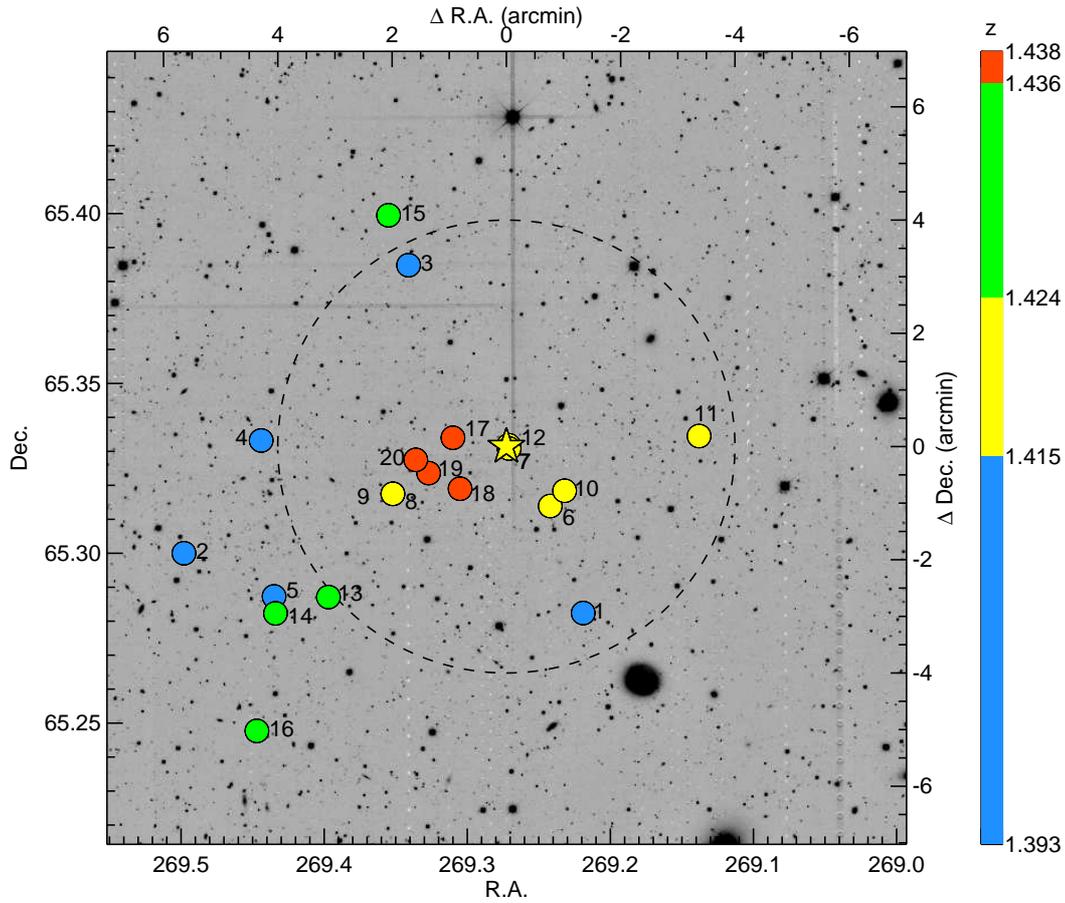}
\end{center}
\caption[The $20$ spectroscopically confirmed members of the structure around 7C~1756+6520.]
{The $20$ spectroscopically confirmed members of the structure around 7C~1756+6520 (plotted on our $J$-band 
data in filled symbols). The four colours (blue, yellow, green and orange,
see redshift scale on the right) indicate different ranges of redshifts: $z<1.415$, $1.415\le z\le1.424$, $1.424<z<1.436$ 
and $z\ge1.436$ respectively.
ID numbers from Table~\ref{members} are also provided for each source.
The dashed circle indicates a distance of $2$~Mpc from the radio galaxy.}
\label{radec2}
\end{figure*}

The spatial distribution of the $20$ objects is shown in Fig.~\ref{radec2}
(the $17$ red symbols in Fig.~\ref{radec} and the three additional
serendipitous objects).  Seven galaxies (IDs $6-12$) have velocities
within $1000$\, km\, s$^{-1}$ of the radio galaxy redshift (i.e.,~the
assumed cluster redshift) and are found within $2$~Mpc of the HzRG
(yellow circles; $1.415\le z \le 1.424$). The lowest redshift sources
(IDs $1-5$; blue circles; $z<1.415$) are found with projected
separations more than $1$~Mpc with respect to the radio galaxy
($3/5$ beyond $2$~Mpc).  Objects with $1.424<z<1.436$ (IDs $13-16$;
green circles) are all more than $2$~Mpc distant from the radio galaxy.
In contrast, the highest redshift sources (IDs $17-20$; orange
circles; $z\ge1.436$) are found within $1\arcmin$ from each other
and appear to form a compact sub-structure.  We therefore
confirm that 7C~1756+6520 is part of a large scale structure
of galaxies composed of (at least) two main galaxy groups --- the
galaxy cluster centered on the radio galaxy redshift ($z=1.4156$)
and a compact galaxy sub-group at $z\sim1.437$. 
Following the methodology of \citet{Beers1990} for a small number of galaxy cluster members, 
we estimated velocity dispersions using the `jackknife of the gapper' estimator. We found that the
velocity dispersion of sources within $1.5$~Mpc from the radio galaxy (i.e.,~ IDs $6-10$ 
+ ID $12$ + IDs $17-20$ + the radio galaxy) is $\Delta v\sim1270\pm180$ km s$^{-1}$. A
velocity dispersion of $\Delta v\sim430\pm205$ km s$^{-1}$ was derived for the galaxy cluster associated 
with the radio galaxy (yellow circles in Fig.~\ref{radec2}). Given the small number of 
galaxies used in the calculation, the velocity dispersions given here are only indicative.

\subsection{A high fraction of AGN cluster members}

Four mid-infrared AGN candidates have been confirmed to be associated with 7C1756+6520 (IDs
$1$, $7$, $10$ and $18$).  Galaxies $4$ and $13$, both targeted as
s$BzK$ galaxies, also show AGN signatures in their spectra with 
strong, broad MgII emission lines.  Assuming that the mid-infrared selected sources are indeed
all active, six AGN (seven with the radio galaxy) have
therefore been spectroscopically confirmed in close proximity both
spatially and in redshift space. Three of them are found within
$1.5\arcmin$ of the radio galaxy. Studying the surface density of
luminous AGN associated with a sample of $330$ galaxy clusters at
$0<z<1.5$, \citet{Galametz2009A} found an excess of AGN within
$0.5$~Mpc of the center of clusters at $z>0.5$. They identify AGN
using three different selection methods including the \citet{Stern2005}
mid-infrared selection. Our spectroscopy brings additional evidence
that galaxy clusters at $z>1$ have a significant fraction of members
being AGN and that they lie preferentially near the cluster center.

\section{Conclusions}

The observations presented here confirm the existence of a large
scale structure of galaxies associated with 7C~1756+6520. We assign
a new reliable redshift of $z=1.4156$ to the radio galaxy.
Our optical spectroscopy demonstrates the efficiency of the $BzK$
selection technique of \citet{Daddi2004} at finding galaxies at
$z>1.4$. It also shows that the modified p$BzK$* criterion presented
in G09 increases the completeness of galaxies at $z>1.4$.  We find twenty
galaxies with spectroscopic redshifts consistent with the
redshift of 7C~1756+6520.  Seven of these galaxies have velocity offsets
$\Delta v < 1000$\, km\, s$^{-1}$ relative to the redshift of the radio
galaxy (assumed here to be the galaxy structure mean redshift) and
are within $2$~Mpc of the radio galaxy.  
A second compact group of four galaxies (all within $1\arcmin$ of each other)
lies at $z\sim1.437$ and forms an associated sub-group
of galaxies, offset by about $1.5\arcmin$ to the east of the radio
galaxy. 7C~1756+6520 is therefore part of a large scale galaxy structure composed
of (at least) two main groups --- a galaxy cluster centered on the radio
galaxy at $z=1.4156$ (with $8$ spectroscopically confirmed members including the radio galaxy)
and a compact group at $z\sim1.437$, $1.5\arcmin$ east of the radio galaxy (with $4$ confirmed members so far).

\begin{acknowledgements}
This work is based on a spectroscopic campaign at the W. M. Keck
Observatory, a scientific partnership between the University of
California and the California Institute of Technology, made possible
by a generous gift of the W. M. Keck Foundation.  We are very
grateful to Tadayuki Kodama for having provided the models of red
sequence presented in this paper. We thank the anonymous referee 
for his/her careful reading of the manuscript and constructive comments. 
The work of DS and RLG was carried
out at Jet Propulsion Laboratory, California Institute of Technology,
under a contract with NASA.  SAS's work was performed under the
auspices of the U.S. Department of Energy, National Nuclear Security
Administration by the University of California, Lawrence Livermore
National Laboratory under contract No. W-7405-Eng-48.

\end{acknowledgements}

\begin{table*}
\caption{Spectroscopic members of the structure around 7C~1756+6520}
\label{members}
\centering
\begin{tabular}{l l l l c l c c c l}
ID			& Name$^{\mathrm{a}}$	& R.A.		& Dec.		& $z_{\rm spec}$ & Q$^{\mathrm{b}}$ &$B^{\mathrm{c}}$ & $z^{\mathrm{c}}$ &   $K^{\mathrm{c}}$	&	Notes	\\ 
			& 			& (J2000)	& (J2000)	&		 &	&	(AB)	&	(AB)	&	(AB)	&               \\ 
\hline
Cl~1756.1				&  AGN.1110	& 17:56:52.56	& 65:16:56.65		&	$1.3935\pm0.0012$	 & B	&	$23.21$	&	$22.14$	&	$20.89$	&	B$2900$, B$3260$, [OII]		\\ 
Cl~1756.2				&  s$BzK$.6355	& 17:57:59.55	& 65:18:00.64	&	$1.4020\pm0.0001$	 & A	&	$24.36$	&	$23.34$	&	$21.54$	&	[OII] 				\\  
Cl~1756.3				&  s$BzK$.9622	& 17:57:21.75	& 65:23:05.04	&	$1.4064\pm0.0005$	 & A	&	$24.00$	&	$22.78$	&	$20.38$	&	[OII] 				\\  
Cl~1756.4				&  s$BzK$.7556	& 17:57:46.54	& 65:20:00.48	&	$1.4081\pm0.0007$	 & A	&	$20.97$	&	$20.77$	&	$20.35$	&	AGN: broad MgII, [OII] 		\\ 
Cl~1756.5				&  p$BzK$.5858	& 17:57:44.40	& 65:17:14.30	&	$1.4110\pm0.0010$	 & B	&	$>27.1$	&	$23.80$	&	$21.62$	&	[OII]; ID$31$ in G09		\\  
Cl~1756.6				&  serendip.1	& 17:56:57.67	& 65:18:49.45		&	$1.4150\pm0.0005$	 & A	&	-	&	-	&	-	&	[OII]				\\ 	
Cl~1756.7$^{\mathrm{d}}$	&  AGN.1354	& 17:57:04.98	& 65:19:51.00		&	$1.4153\pm0.0003$	 & B	&	$26.34$	&	$22.03$	&	-	&	AGN: broad MgII			\\  
HzRG				&  7C~1756+6520	& 17:57:05.48	& 65:19:53.75	&	$1.4156\pm0.0001$	 & A	&	$>27.1$	&	$21.40$	&	$20.17$	&	AGN: [NeV], [OII], [NeIII]	\\ 
Cl~1756.8				&  serendip.2	& 17:57:25.00	& 65:19:04.83		&	$1.4157\pm0.0010$	 & A	&	$26.32$	&	$24.34$	&	$>23.4$	&	[OII]				\\  
Cl~1756.9				&  s$BzK$.6997	& 17:57:24.43	& 65:19:03.87	&       $1.4157\pm0.0006$	 & A	&	$25.21$	&	$24.21$	&	$22.56$	&	[OII]				\\  
Cl~1756.10			&  AGN.1317	& 17:56:55.75	& 65:19:07.00		&	$1.4162\pm0.0005$	 & A	&	$20.16$	&	$19.46$	&	$19.01$	&	AGN: broad MgII, [NeV], [OII]	\\  
Cl~1756.11			&  s$BzK$.7624	& 17:56:33.16	& 65:20:04.46	&	$1.4236\pm0.0001$	 & A	&	$22.51$	&	$22.39$	&	$21.30$	&	FeII+MgII absn, [OII	]	\\  
Cl~1756.12$^{\mathrm{e}}$	&  p$BzK$.7523	& 17:57:05.04	& 65:19:54.50	&	$1.4244\pm0.0004$	 & A	&	$>27.1$	&	$23.47$	&	$21.31$	&	[OII]; ID$43$ in G09		\\  
Cl~1756.13			&  s$BzK$.5860	& 17:57:35.34	& 65:17:14.39	&	$1.4268\pm0.0005$	 & A	&	$22.96$	&	$22.35$	&	$21.31$	&	AGN: broad MgII, [OII]		\\  
Cl~1756.14			&  s$BzK$.5699	& 17:57:44.06	& 65:16:57.11	&	$1.4274\pm0.0004$	 & A	&	$25.89$	&	$24.20$	&	$22.71$	&	[OII]				\\  
Cl~1756.15			&  p$BzK$.10235	& 17:57:25.20	& 65:23:58.19	&	$1.4277\pm0.0010$	 & B	&	$>27.1$	&	$23.45$	&	$21.76$	&	[OII]; ID$63$ in G09	\\
Cl~1756.16			&  s$BzK$.4449	& 17:57:47.40	& 65:14:52.17	&	$1.4326\pm0.0002$	 & A	&	$24.93$	&	$23.89$	&	$23.02$	&	[OII]				\\ 
Cl~1756.17			&  s$BzK$.7625	& 17:57:14.41	& 65:20:02.40	&	$1.4366\pm0.0001$	 & A	&	$>27.1$	&	$24.46$	&	$22.95$	&	[OII]				\\ 
Cl~1756.18			&  AGN.1206	& 17:57:13.08	& 65:19:08.37		&	$1.4371\pm0.0002$	 & A	&	$>27.1$	&	$>25.0$	&	$21.49$	&	[OII]				\\  
Cl~1756.19			&  s$BzK$.7208	& 17:57:18.31	& 65:19:24.94	&	$1.4374\pm0.0002$	 & A	&	$24.93$	&	$23.46$	&	$22.00$	&	[OII]				\\  
Cl~1756.20			&  serendip.3	& 17:57:20.76	& 65:19:39.14		&	$1.4379\pm0.0007$	 & A	&	$25.29$	&	$23.18$	&	$21.50$	&	[OII]				\\ 
\hline
\end{tabular} 
\begin{list}{}{}
\item[$^{\mathrm{a}}$] The names are composed of the selection
technique for the source --- AGN, s$BzK$ or p$BzK$ ($\equiv$ p$BzK$*
here) --- followed by the identification number in our $K$-band
($3.6\mu$m) catalogue for $BzK$ (AGN) targets.  Sources serendipitously
identified are given a simple 'serendip.\#' designation.

\item[$^{\mathrm{b}}$] Q indicates the quality of the redshift,
either `A' indicating a highly certain redshift, or `B' indicating
a high level of confidence (see \S3.2).

\item[$^{\mathrm{c}}$] Magnitudes were derived using SExtractor
MAG\_AUTO and are therefore slightly different from the aperture
magnitudes used to derive colours for candidate selection.


\item[$^{\mathrm{d}}$] AGN found within $6\arcsec$ of 7C~1756+6520.
No accurate magnitude is available in $K$ due to an artifact caused by a
bright star in the near-infrared data. See Fig.~12 in G09.

\item[$^{\mathrm{e}}$] p$BzK$* galaxy found within $3\arcsec$ of
7C~1756+6520.

\end{list}
\end{table*}




\begin{table*}
\caption{Other spectroscopic redshifts}
\label{others}
\centering
\begin{tabular}{l l l c l c c c l}
Name$^{\mathrm{a}}$ 	& R.A.		& Dec.		&$z_{\rm spec}$&Q$^{\mathrm{b}}$&$B^{\mathrm{c}}$&	$z^{\mathrm{c}}$	&	$K^{\mathrm{c}}$	&	Notes	\\
			& (J2000)	& (J2000)	&		&		&	(AB)	&	(AB)	&	(AB)	&		\\
\hline
s$BzK$.8902		& 17:57:15.84	& 65:22:04.07	&	$0.3347\pm0.0005$	&	B	&	$24.28$	&	$22.93$	&	$21.88$	&	[OIII]				\\  
AGN.1005		& 17:57:05.48	& 65:19:53.75	&	$0.3658\pm0.0001$	&	A	&	$22.25$	&	$19.51$	&	$18.78$	&	CaHK, H$\beta$, H$\alpha$, [NII]		\\  
s$BzK$.4858		& 17:57:48.87	& 65:15:31.59	&	$0.4265\pm0.0001$	&	A	&	$24.18$	&	$22.85$	&	$22.14$	&	H$\beta$, [OIII], H$\alpha$		\\  	
s$BzK$.9364		& 17:57:29.28	& 65:22:42.61	&	$0.5318\pm0.0008$	&	B	&	$>27.1$	&	$23.62$	&	$21.06$	&	[OIII]				\\ 
s$BzK$.7895		& 17:57:05.04	& 65:20:31.57	&	$0.7529\pm0.0003$	&	A	&	$22.70$	&	$21.75$	&	$20.94$	&	[OII], H$\beta$		\\ 	
s$BzK$.6858		& 17:56:58.08	& 65:18:50.05	&	$0.7533\pm0.0004$	&	B	&	$23.06$	&	$23.11$	&	$21.02$	&	[OII]				\\  
s$BzK$.7921		& 17:56:39.84	& 65:20:30.52	&	$1.0196\pm0.0001$	&	A	&	$22.73$	&	$23.67$	&	$22.79$	&	[OII], [NeIII]		\\  
s$BzK$.11226		& 17:57:19.44	& 65:25:24.60	&	$1.0505\pm0.0002$	&	A	&	$24.45$	&	$23.83$	&	$23.19$	&	[OII], [NeIII], H$\beta$\\  	
s$BzK$.9844		& 17:57:16.38  	& 65:23:19.85	&	$1.1227\pm0.0005$	&	A	&	$26.00$	&	$24.18$	&	$21.99$	&	[OII]				\\  
p$BzK$.7866		& 17:56:42.72	& 65:20:27.97	&	$1.1825\pm0.0010$	&	B	&	$>27.1$	&	$23.11$	&	$20.93$	&	[OII]			\\ 
AGN.1767		& 17:57:15.33	& 65:21:56.14	&	$1.1827\pm0.0004$	&	A	&	$>27.1$	&	$>25.0$	&	$19.58$	&	MgII, [OII]		\\  
s$BzK$.9430		& 17:55:53.37	& 65:22:48.02	&	$1.2074\pm0.0001$	&	A	&	$23.44$	&	$22.79$	&	$21.73$	&	AGN: MgII absn, [OII], [NeIII], H$\zeta$, H$\delta$, \\  
p$BzK$.7359		& 17:57:12.44	& 65:19:45.10	&	$1.2122\pm0.0002$	&	A	&	$26.65$	&	$22.63$	&	$20.18$	&	[OII]			\\  
p$BzK$.4948		& 17:57:16.54	& 65:15:44.17 	&	$1.2130\pm0.0003$	&	A	&	$>27.1$	&	$22.79$	&	$20.49$	&	[OII]			\\  	 
s$BzK$.11168		& 17:56:57.97  	& 65:25:20.26	&	$1.2359\pm0.0001$	&	A	&	$25.46$	&	$23.93$	&	$22.37$	&	[OII]			\\  
s$BzK$.12407		& 17:56:56.58  	& 65:26:36.42	&	$1.2461\pm0.0001$	&	A	&	$25.02$	&	$24.33$	&	$22.99$	&	[OII]			\\  	
s$BzK$.8631		& 17:56:31.31	& 65:21:41.17	&	$1.2586\pm0.0002$	&	A	&	$23.90$	&	$23.06$	&	$21.36$	& 	[OII]				\\  	
p$BzK$.10254		& 17:57:11.04	& 65:23:56.76	&	$1.2598\pm0.0001$	&	A	&	$>27.1$	&	$23.40$	&	$21.20$	&	[NeV], [OII], [NeIII]	\\  
s$BzK$.2137		& 17:56:52.56	& 65:11:09.95	&	$1.2830\pm0.0001$	&	A	&	$24.09$	&	$23.96$	&	$22.61$	&	H$\beta$, [OIII], [NII]	\\  
s$BzK$.8410		& 17:57:25.15  	& 65:21:18.01	&	$1.4564\pm0.0003$	&	A	&	$25.37$	&	$23.75$	&	$21.90$	&	[OII]				\\  
s$BzK$.7159		& 17:57:28.56	& 65:19:17.93	&	$1.4576\pm0.0002$	&	A	&	$24.70$	&	$23.05$	&	$22.44$	&	[OII]				\\  	
s$BzK$.6886		& 17:57:37.75  	& 65:18:53.95	&	$1.4663\pm0.0004$	&	A	&	$25.82$	&	$23.25$	&	$21.70$	&	[OII]				\\  
s$BzK$.8761		& 17:56:57.03	& 65:21:53.86	&	$1.4764\pm0.0003$	&	A	&	$23.35$	&	$22.83$	&	$20.37$	&	[OII]				\\  	
s$BzK$.8799		& 17:57:11.43  	& 65:21:53.94	&	$1.5589\pm0.0001$	&	A	&	$23.59$	&	$22.26$	&	$21.14$	&	CIII], MgII abns, [OII]	\\  	
s$BzK$.2998		& 17:56:56.40	& 65:12:32.76	&	$1.5620\pm0.0004$	&	A	&	$24.19$	&	$23.79$	&	$22.88$	&	[OII]				\\  	
s$BzK$.9006		& 17:56:28.73	& 65:22:10.94	&	$1.5943\pm0.0002$	&	A	&	$22.95$	&	$22.84$	&	$22.27$	&	[OII]				\\  
s$BzK$.9788		& 17:55:49.35	& 65:23:16.14	&	$1.595\pm0.001$	&	A	&	$23.93$	&	$22.98$	&	$22.18$	&	FeII+MgII absn	\\  	
s$BzK$.11725		& 17:56:30.31  	& 65:25:31.69	&	$1.725\pm0.005$	&	B	&	$22.35$	&	$21.78$	&	$21.00$	&	AGN: broad MgII			\\  
s$BzK$.5102		& 17:57:57.61	& 65:15:55.59	&	$1.979\pm0.002$	&	A	&	$23.59$	&	$22.84$	&	$22.21$	&	AlII, MgII			\\ 	
s$BzK$.6330		& 17:57:37.08	& 65:17:58.83	&	$2.218\pm0.001$	&	A	&	$23.82$	&	$23.23$	&	$21.94$	&	AGN: CIV, CIII], MgII	\\  
s$BzK$.3534		& 17:58:01.74	& 65:13:31.73	&	$2.256\pm0.002$	&	A	&	$23.73$	&	$22.84$	&	$21.82$	&	CIV+AlII absn			\\	
AGN.1748		& 17:57:20.07  	& 65:21:56.96	&	$2.399\pm0.001$	&	A	&	$23.06$	&	$21.39$	&	$20.58$	&	Quasar: CIV, CIII]	\\  
AGN.1340		& 17:57:30.88	& 65:21:22.32	&	$2.402\pm0.001$	&	A	&	$24.05$	&	$21.80$	&	$19.44$	&	Quasar: CIV, CIII]	\\  
s$BzK$.3718		& 17:58:01.19	& 65:13:49.25	&	$2.4020\pm0.0005$	&	A	&	$22.65$	&	$21.75$	&	$21.76$	&	CIV+AlII absn			\\	
AGN.1213		& 17:57:33.22	& 65:20:26.24	&	$3.702\pm0.002$	&	A	&	$22.73$	&	$19.72$	&	$18.95$	&	Quasar: Ly$\alpha$, CIV, CIII]	\\  
\hline
\end{tabular} 
\begin{list}{}{}
\item[$^{\mathrm{a}}$] The names are composed of the selection
technique for the source --- AGN, s$BzK$ or p$BzK$ ($\equiv$ p$BzK$*
here) --- followed by the identification number in our $K$-band
($3.6\mu$m) catalogue for $BzK$ (AGN) targets.  

\item[$^{\mathrm{b}}$] Q indicates the quality of the redshift,
either `A' indicating a highly certain redshift, or `B' indicating
a high level of confidence (see \S3.2).

\item[$^{\mathrm{c}}$] Magnitudes were derived using SExtractor
MAG\_AUTO and are therefore slightly different from the aperture
magnitudes used to derive colours for candidate selection.

\end{list}
\end{table*}


\newpage

\newcommand{\noopsort}[1]{}

\end{document}